\documentclass[symmetry,article,submit,oneauthor,pdftex]{Definitions/mdpi}
\firstpage{1}
\pubvolume{xx}
\issuenum{1}
\articlenumber{5}
\pubyear{2020}
\copyrightyear{2020}

\newcommand{\ie}{\textit{i.e.}}
\newcommand{\eg}{\textit{e.g.}}

\pdfoutput=1

\usepackage{aas_macros}
\usepackage{acronym}
\usepackage{amsmath,amsfonts,amssymb}

\usepackage{tensind}
\tensordelimiter{?}
\tensorformat{lrc}

\Title{Binary Neutron Star Merger Simulations with a Calibrated Turbulence Model}

\Author{David Radice$^{1,2,3}$\orcidA{}}

\AuthorNames{David Radice}

\address{%
$^{1}$ \quad Institute for Gravitation \& the Cosmos, The Pennsylvania State University, University Park, PA 16802\\
$^{2}$ \quad Department of Physics, The Pennsylvania State University, University Park, PA 16802\\
$^{3}$ \quad Department of Astronomy \& Astrophysics, The Pennsylvania State University, University Park, PA 16802}

\corres{Correspondence: david.radice@psu.edu}

\abstract{Magnetohydrodynamic (MHD) turbulence in neutron star (NS)
merger remnants can impact their evolution and multimessenger
signatures, complicating the interpretation of present and future
observations. Due to the high Reynolds numbers and the large
computational costs of numerical relativity simulations, resolving all
the relevant scales of the turbulence will be impossible for the
foreseeable future. Here, we adopt a method to include subgrid-scale
turbulence in moderate resolution simulations by extending the
large-eddy simulation (LES) method to general relativity (GR). We
calibrate our subgrid turbulence model with results from
very-high-resolution GRMHD simulations, and we use it to perform NS
merger simulations and study the impact of turbulence. We find that
turbulence has a quantitative, but not qualitative impact on the
evolution of NS merger remnants, on their gravitational wave signatures,
and on the outflows generated in binary NS mergers. Our approach
provides a viable path to quantify uncertainties due to turbulence in NS
mergers.}


\begin{document}

\section{Introduction}
Binary neutron star (BNS)\acused{BNS} mergers are prime targets for the
ground-based laser interferometric gravitational-wave (GW)\acused{GW}
detectors LIGO \cite{TheLIGOScientific:2014jea}, Virgo
\cite{TheVirgo:2014hva}, and KAGRA \cite{Aso:2013eba}. \ac{BNS} mergers
generate loud \ac{GW} signals and can also power bright \ac{EM}
transients \cite{Eichler:1989ve, Narayan:1992iy, Berger:2013jza,
Kumar:2014upa, Fernandez:2015use, Metzger:2019zeh, Nakar:2019fza}, as
demonstrated by the extraordinary multimessenger observations of
GW170817 \cite{TheLIGOScientific:2017qsa, Abbott:2018wiz}. Finally,
\ac{BNS} mergers can eject neutron rich material which subsequently
produces heavy elements, such as gold and uranium, through the r-process
\cite{Eichler:1989ve, Wanajo:2014wha, Hotokezaka:2018aui,
Cowan:2019pkx}. At the time of writing, one more \ac{BNS} \ac{GW} event
after GW170817 has been announced by the \ac{LVC}: GW190425
\cite{Abbott:2020uma, Foley:2020kus}. However, several more candidates
have been reported and are currently being analyzed by the \ac{LVC}
\cite{gracedb}. Many more detections are expected in the next years as
\ac{GW} observatories improve their sensitives and as more facilities
are added to the global network of detectors \cite{Aasi:2013wya}.

Multimessenger observations of \ac{BNS} mergers are starting to
constrain the poorly known properties of matter at extreme densities
\cite{Hinderer:2007mb, Damour:2009vw, Damour:2012yf,
TheLIGOScientific:2017qsa, Margalit:2017dij, Abbott:2018wiz,
Annala:2017llu, De:2018uhw, Ruiz:2017due, Bauswein:2017vtn,
Radice:2017lry, Most:2018hfd, Tews:2018iwm, Radice:2018ozg,
Kiuchi:2019lls, Shibata:2019ctb, Capano:2019eae, Annala:2019puf,
Dietrich:2020lps} and the physical processes powering \acp{SGRB}
\cite{Ruiz:2016rai, Monitor:2017mdv, Lazzati:2017zsj, Xie:2018vya,
Hajela:2019mjy, Lazzati:2020vbo}. They are also beginning to reveal the
role played by compact binary mergers in the chemical enrichment of the
galaxy with r-process elements \cite{Rosswog:1998hy, Lee:2009uc,
Hotokezaka:2012ze, Bauswein:2013yna, Wanajo:2014wha, Perego:2014fma,
Fernandez:2015use, Sekiguchi:2016bjd, Foucart:2015gaa, Radice:2016dwd,
Lehner:2016lxy, Dietrich:2016hky, Siegel:2017nub, Kasen:2017sxr,
Foucart:2016rxm, Fujibayashi:2017puw, Fernandez:2018kax, Radice:2018pdn,
Miller:2019dpt, Vincent:2019kor, Nedora:2019jhl, Fujibayashi:2020qda}.
The key to the solution of some of the most pressing open problems in
nuclear and high-energy astrophysics -- such as the origin of heavy
elements, the nature of \acp{NS}, and the origin of \acp{SGRB} -- is
encoded in these and future observations. However, theory is essential
to turn observations into answers.

Numerical relativity (NR)\acused{NR} simulations are the only tool able
to study the dynamics of \ac{BNS} mergers in the strong field regime and
its connection to the multimessenger signals they produce.
State-of-the-art \ac{NR} simulations include a microphysical treatment
of dense matter, the impact of weak reactions and neutrino radiation,
and magnetic effects \cite{Palenzuela:2015dqa, Most:2019kfe,
Mosta:2020hlh, Jesse:2020oss}. Even though modern simulations ostensibly
include all of the physics believed to determine the outcome of \ac{BNS}
mergers, the long-term evolution of binaries after merger remains poorly
known, \eg, \cite{Radice:2018xqa}.  Leading sources of uncertainty are
connected to our limited knowledge of the behavior of matter at extreme
densities and temperatures, the crudeness with which neutrino radiation is
treated in the simulations, and our inability to simulate these systems
at sufficiently high resolution to resolve the turbulent cascade and for
sufficiently long times \cite{Radice:2020ddv}. This work is part of our
ongoing effort to address this last issue. 

It is known that the matter flow after merger is subject to a number of
\ac{MHD}\acused{GRMHD}\acused{GRHD} instabilities, such as the \ac{KH}
instability and the \ac{MRI} \cite{Obergaulinger:2010gf,
Bucciantini:2011kx, Siegel:2013nrw, Kiuchi:2014hja, Giacomazzo:2014qba,
Kiuchi:2015sga, Kiuchi:2017zzg}. These inject turbulence at very small
scale and can potentially impact the qualitative outcome of the merger
\cite{Duez:2004nf, Duez:2006qe, Hotokezaka:2013iia, Ciolfi:2019fie}.
However, even the best resolved \ac{GRMHD} simulations to date
\cite{Kiuchi:2015sga, Kiuchi:2017zzg, Mosta:2020hlh} cannot capture the
scale of the fastest growing mode of the \ac{MRI}, unless artificially
large initial magnetic fields are adopted to increase the cutoff length
scales associated with some of these instabilities. Even in these cases,
simulations are far from being able to capture the dynamics of the
turbulent cascade all the way to the viscous scale, at which neutrino
viscosity and drag damps the turbulent eddies \cite{Guilet:2016sqd}, as
would be required for a DNS simulation.

In Ref.~\cite{Radice:2017zta} we proposed the
\ac{GRLES}\acused{LES}\acused{GR} method as an alternative to performing
ultra-high resolution \ac{GRMHD} simulations.  In particular, we
proposed to evolve the coarse-grained \ac{GRHD} equations with a
turbulent closure models design to capture the effect of turbulence
operating at sub-grid scales. In parallel, a similar, but technically
distinct, approach based on the Israel-Stewart formalism was proposed by
Shibata and collaborators \cite{Shibata:2017jyf}. More recently, a
rigorous first-principle theory of relativistic turbulence that, among
other things, strengthens the mathematical foundation for the \ac{GRLES}
method, has been proposed by Eyink and Drivas~\cite{Eyink:2017zfz}. An
extension of the method to \ac{GRMHD}, taking into account also terms
that we neglected in our initial formulation (more on this below), has
been proposed in Refs.~\cite{Carrasco:2019uzl, Vigano:2020ouc}. Rosofsky
and Huerta \cite{Rosofsky:2020zsl} proposed to use machine learning to
calibrate subgrid turbulence models for 2D MHD. Finally, a variant of
the \ac{GRLES} method has also been implemented into the SpEC code by
the SXS collaboration to perform 2D axisymmetric
simulations~\cite{Jesse:2020oss}.

The \ac{GRLES} or viscous approaches are the only way to perform
long-term simulations of the postmerger evolution for multiple systems
\cite{Fujibayashi:2017puw, Nedora:2019jhl, Fujibayashi:2020qda}.
However, the results from these simulations inevitably depend on the
adopted subgrid model. In earlier work we used turbulence models based
on dimensional analysis and linear perturbation theory. Here, we
calibrate a subgrid model using results from very high resolution
\ac{GRMHD} simulations performed by Kiuchi and
collaborators~\cite{Kiuchi:2017zzg}, which were able to resolve all the
unstable scales of the \ac{MRI} for a binary system with an initially
large magnetic field. We perform \ac{BNS} merger simulations with
microphysics and compare results obtained with the newly calibrated
turbulence model with those obtained using the prescription we proposed
in Ref.~\cite{Radice:2017zta}, which was used in several other works
\cite{Zappa:2017xba, Radice:2018pdn, Radice:2018ghv, Perego:2019adq,
Nedora:2019jhl, Bernuzzi:2020txg}, and to those obtained with
traditional \ac{GRHD} simulations having no subgrid model. We find that
turbulence can impact the postmerger evolution of \acp{BNS} in a
quantitative way, and we discuss the implications for the interpretation
of synthetic \ac{GW}, \ac{EM}, and nucleosynthesis yields from \ac{BNS}
merger simulations.

The rest of this paper is organized as follows. In
Section~\ref{sec:methods} we review the \ac{GRLES} formalism and discuss
the calibration of the subgrid model. In Section \ref{sec:results} we
present our simulation results. Finally, Section \ref{sec:conclusions}
is dedicated to discussion and conclusions.

\section{Methods}
\label{sec:methods}

\subsection{WhiskyTHC}
All simulations are performed with the \texttt{WhiskyTHC} code
\cite{Radice:2012cu, Radice:2013hxh, Radice:2013xpa, Radice:2015nva}.
\texttt{WhiskyTHC} separately evolves the proton and neutron number
densities
\begin{equation}\label{eq:number}
  \nabla_\mu ( J_{p,n}^\mu ) = R_{p,n}\,,
\end{equation}
where $J_{p,n}^\mu = n_{p,n} u^\mu$ are the proton and neutron
four-currents, $n_p = Y_e n$ is the proton number density, $n_n$ is the
neutron number density, $n = n_p + n_n$ is the baryon number density
(including baryons in nuclei), $u^\mu$ the fluid four-velocity, and
$Y_e$ is the electron fraction of the material. $R_p = - R_n$ is the net
lepton number deposition rate due to the absorption and emission of
neutrinos and anti-neutrinos, which is computed using the M0 scheme
\cite{Radice:2016dwd, Radice:2018pdn}.

\ac{NS} matter is treated as a perfect fluid with stress energy tensor
\begin{equation}
  T_{\mu\nu} = (e + p) u_\mu u_\nu + p g_{\mu\nu}\,,
\end{equation}
where $e$ is the energy density and $p$ the pressure. We solve the
equations for the balance of energy and momentum
\begin{equation}\label{eq:euler}
  \nabla_\nu T^{\mu\nu} = Q u^\mu\,,
\end{equation}
where $Q$ is the net energy deposition rate due to the absorption
and emission of neutrinos, also treated using the M0 scheme.

The spacetime is evolved using the Z4c formulation of Einstein's
equations \cite{Bernuzzi:2009ex, Hilditch:2012fp} as implemented in the
\texttt{CTGamma} code \cite{Pollney:2009yz, Reisswig:2013sqa}, which is
part of the \texttt{Einstein Toolkit} \cite{EinsteinToolkit,
Loffler:2011ay}. \texttt{CTGamma} and \texttt{WhiskyTHC} are coupled
using the method of lines. For this work we use the optimal
strongly-stability preserving third-order Runge-Kutta scheme
\citep{Gottlieb:2008a} as time integrator. Mesh adaptivity is handled
using the \texttt{Carpet} mesh driver \cite{Schnetter:2003rb} which
implements Berger-Oliger style adaptive mesh refinement (AMR) with
subcycling in time and refluxing \cite{Berger:1984zza,
1989JCoPh..82...64B, Reisswig:2012nc}.

\subsection{GRLES}
\label{sec:grles}
According to the Valencia formalism for \ac{GRHD} \cite{Banyuls:1997zz}
the fluid for velocity is decomposed as the sum of a vector parallel and one
orthogonal to the $t = \mathrm{const}$ hypersurface normal $n^\mu$ (not to be
confused with the neutron and proton number densities) as:
\begin{equation}
  u^\mu = (- u_\mu n^\mu) (n^\mu + v^\mu) =: W (n^\mu + v^\mu)\,,
\end{equation}
where $W$ is the Lorentz factor and $v^\mu$ is the three velocity.
Accordingly, the proton and neutron currents can be written as 
\begin{equation}
  J_{n,p}^\mu =  n_{n,p} W (n^\mu + v^\mu) =: D_{n,p} (n^\mu + v^\mu)\,.
\end{equation}
In a similar way, the stress energy tensor is decomposed as
\begin{equation}\label{eq:tmunu.decomp}
  ?[c]T_\mu_\nu? = E n_\mu n_\nu + S_{\mu} n_{\nu} + S_{\nu} n_{\mu} + S_{\mu\nu}\,,
\end{equation}
where
\begin{align}
  &E = T_{\mu\nu} n^\mu n^\nu = (e + p) W^2 -p\,,  \\
  &S_\mu = - \gamma_{\mu\alpha} n_\beta T^{\alpha\beta} = (e + p) W^2
  v_\mu\,, \\
  &S_{\mu\nu} = \gamma_{\mu\alpha} \gamma_{\nu\beta} T^{\alpha\beta}
  = S_\mu v_\nu + p \gamma_{\mu\nu}\,,
\end{align}
are respectively the energy density, the linear momentum density, and
the stress tensor in a frame having four-velocity $n^\mu$, and
$\gamma_{\mu\nu}$ is the spatial metric.

With these definitions in place, and neglecting the neutrino source
terms to keep the notation simple, the \ac{GRHD} equations read
\begin{align}
  \label{eq:valencia.1}
  \partial_t \big(\sqrt{\gamma} D_{n,p}\big) + \partial_j \Big[
    \alpha\sqrt{\gamma} \big(v^j + n^j) D_{n,p} \Big] &= 0\,, \\
  \label{eq:valencia.2}
  \partial_t \big(\sqrt{\gamma} S_i\big) + \partial_j \Big[
    \alpha\sqrt{\gamma}\big( ?[c]S_i^j? + S_i n^j  \big)
    \Big] &= \alpha\sqrt{\gamma}\Big(\frac{1}{2} S^{jk} \partial_i
    \gamma_{jk} + \frac{1}{\alpha} S_k \partial_i \beta^k - E \partial_i
    \log \alpha \Big)\,, \\
  \label{eq:valencia.3}
  \partial_t \big(\sqrt{\gamma} E\big) + \partial_j \Big[
    \alpha\sqrt{\gamma}\big(S^j + E n^j\big)\Big] &= \alpha\sqrt{\gamma}
    \Big(K_{ij} S^{ij} - S^i \partial_i \log \alpha \Big)\,.
\end{align}

The \ac{GRLES} methodology derives a set of equations for the large
scale dynamics of the flow, in the sense precisely defined in
Ref.~\cite{Eyink:2017zfz}, by applying a linear filtering operator $X
\mapsto \overline{X}$ to derive a set of equations for the coarse
grained quantities. For example, the cell averaging done in the context
of a finite volume method can be considered as a type of filtering. The
averaged
equations read:
\begin{align}
  \label{eq:grles.1}
  \partial_t \big(\sqrt{\gamma} \overline{D_{n,p}}\big) + \partial_j \Big[
    \alpha\sqrt{\gamma} \big(\overline{D_{n,p} v^j} + \overline{D} n^j) \Big] 
    &= 0\,, \\
  \label{eq:grles.2}
  \partial_t \big(\sqrt{\gamma} \overline{S_i}\big) + \partial_j \Big[
    \alpha\sqrt{\gamma}\big( \overline{?[c]S_i^j?} + \overline{S_i} n^j  \big)
    \Big] &= \alpha\sqrt{\gamma}\Big(\frac{1}{2} \overline{S^{jk}} \partial_i
    \gamma_{jk} + \frac{1}{\alpha} \overline{S_k} \partial_i \beta^k -
    \overline{E} \partial_i
    \log \alpha \Big)\,, \\
  \label{eq:grles.3}
  \partial_t \big(\sqrt{\gamma} \overline{E}\big) + \partial_j \Big[
    \alpha\sqrt{\gamma}\big(\overline{S^j} + \overline{E} n^j\big)\Big] 
    &= \alpha\sqrt{\gamma} \Big(K_{ij} \overline{S^{ij}} -
    \overline{S^i} \partial_i \log \alpha \Big)\,.
\end{align}
Here, we have implicitly assumed that the metric quantities are
unaffected by averaging, because they are already large scale
quantities. This is the only approximation made when going from
Eqs.~(\ref{eq:valencia.1})-(\ref{eq:valencia.3}) to
Eqs.~(\ref{eq:grles.1})-(\ref{eq:grles.3}). Although these equations can
be considere exact, they are obviously not closed, since not all terms
can be expressed solely as a function of the evolved quantities
$\overline{D_{n,p}}, \overline{S_i},$ and $\overline{E}$. This is a
manifestation of the nonlinearity of the equations. To close the
equations it is necessary to provide a closure for some of the terms.
The most obvious terms that need to be closed are the quadratic terms:
\begin{align}
  \overline{S_{ij}} = \overline{S_i} \overline{v_j} + \overline{p}
  \delta_{ij} + \tau_{ij}\,, && 
  \overline{D v^i} = \overline{D} \overline{v^i} + \mu^i\,,
\end{align}
The correlation terms $\tau_{ij}$ and $\mu^i$ are the subgrid-scale, or
turbulent, stress and rest-mass diffusion. We remark that these terms
are always present in any numerical discretization of the \ac{GRHD}
equations even if not explicitly included: this is the so-called
implicit large-eddy simulation (ILES) approach.

Since the \ac{EOS} is also non-linear, the filtered pressure is not
equal to the \ac{EOS} evaluated from the coarse-grained
quantities, so an additional closure would also be needed when
evaluating $\overline{p}$, that is:
\begin{equation}
  \overline{p} = p(\overline{D_{n,p}}, \overline{S_i}, \overline{E}) + \Pi
\end{equation}
Similarly, the three velocity $\overline{v^i}$ is also a nonlinear
function of the evolved coarse-grained quantities, so we would need to
include a closure also for $\overline{v^i}$. These terms are treated in
full generality in Refs.~\cite{Carrasco:2019uzl, Vigano:2020ouc}, to
which we refer for the details. Here, we neglect these corrections, \eg,
we assume $\Pi = 0$, because we expect them to be subdominant, since
turbulence in the postmerger remnant is subsonic and subrelativistic,
meaning that its character should be fully captured by $\tau_{ij}$. This
assumption could in principle be verified using \ac{GRMHD} simulation
data. However, the simulations data that we use for calibration
\cite{Kiuchi:2017zzg} is not publicly available, and we only have access
to the value of the $\alpha$ parameter, which maps to
$\tau_{r\phi}$. Consequently, we cannot check the validity of this
assumption.

We employ the relativistic extension of the classical turbulence closure
of Smagorinsky \cite{Smagorinsky:1963a}, which we proposed in
Ref.~\cite{Radice:2017zta}:
\begin{align}\label{eq:turb.visc}
  \tau_{ij} = - 2 \nu_T (e + p) W^2 \left[ \frac{1}{2} \Big( D_{i}
  \overline{v_{j}} + D_{j} \overline{v_i} \Big) -
  \frac{1}{3} D_k \overline{v^k} \gamma_{ij} \right]\,, && \mu^i = 0\,,
\end{align}
where $D_i$ is the covariant derivative associated with $\gamma_{ij}$,
and $\nu_T$, the turbulent viscosity. On the basis of dimensional
analysis arguments it is natural to parametrize $\nu_T$ in terms of a
characteristic velocity, the sound speed $c_s$, and a characteristic
length scale of turbulence, the mixing length $\ell_{\rm mix}$, as
\begin{equation}\label{eq:turb.visc.2}
  \nu_T = \ell_{\rm mix}\, c_s\,.
\end{equation}
For \ac{MRI}-driven turbulence, one can assume $\ell_{\rm mix}$ to be
related to the length scale of the most unstable mode of the MRI
\cite{Duez:2006qe}
\begin{equation}\label{eq:lambdamri}
  \lambda_{\rm MRI} \sim 20\ {\rm m}\ \left(\frac{\Omega}{6\ {\rm rad}\ {\rm
  ms}^{-1}}\right)^{-1} \left(\frac{B}{10^{15}\ {\rm G}}\right)\,,
\end{equation}
which is the scale at which turbulence is predominantly driven 
according to linear theory. Accordingly, in our previous work we
explored the impact of turbulence by varying $\ell_{\rm mix}$ between
$0$ and $50\ {\rm m}$ respectively corresponding to no and very
efficient turbulent mixing.

\begin{figure}
  \centering
  \includegraphics[width=5in]{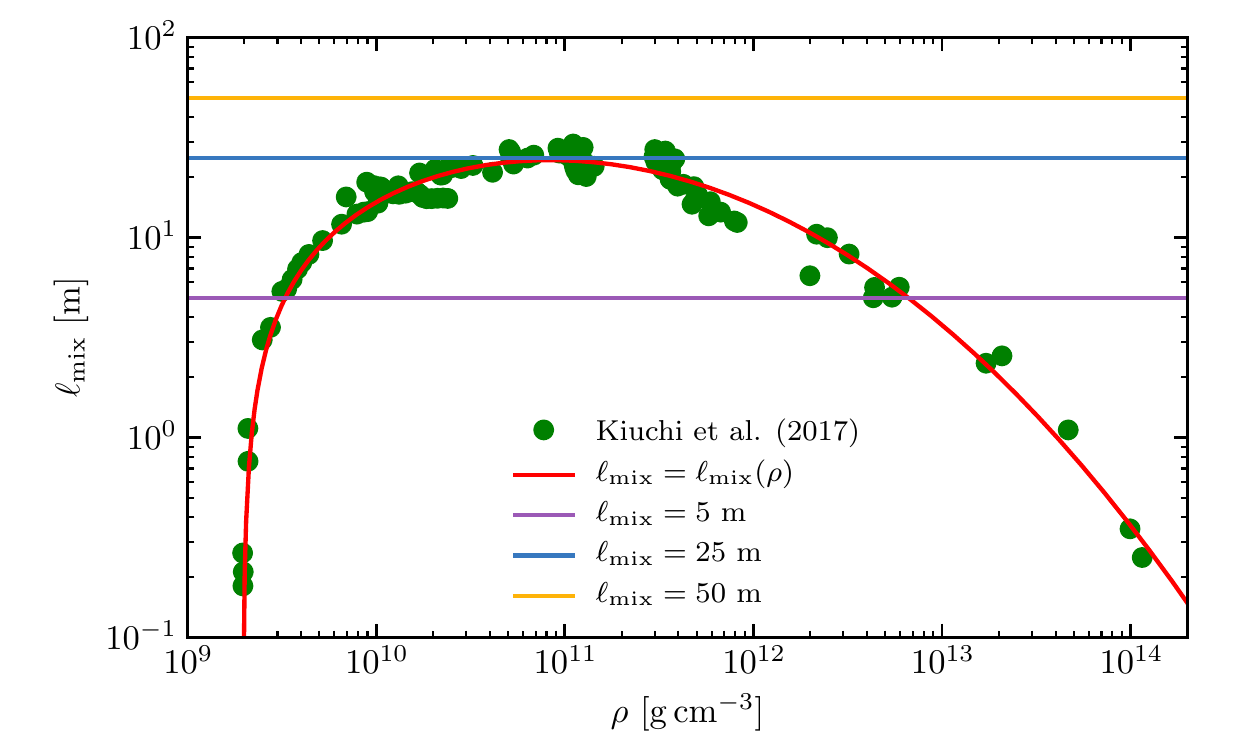}
  \caption{Mixing length evaluated from the GRMHD simulations of Kiuchi
  et al. \cite{Kiuchi:2017zzg} (dots) and fit employed in this work (red
  line). Also shown are the other values of $\ell_{\rm mix}$ employed in
  this and in our previous works. GRMHD simulations favor relatively
  small values $\ell_{\rm mix}$ in agreement with the simple analytic
  estimate in Eq.~\eqref{eq:lambdamri}.}
  \label{fig:lmix_fit}
\end{figure}

In the context of accretion disk theory, turbulent viscosity is typically
parametrized in terms of a dimensionless constant $\alpha$ linked to
$\ell_{\rm mix}$ through the relation $\ell_{\rm mix} = \alpha\, c_s\,
\Omega^{-1}$, where $\Omega$ is the angular velocity of the fluid
\cite{1973A&A....24..337S}. Recently, Kiuchi and
collaborators~\cite{Kiuchi:2017zzg} performed very high resolution
\ac{GRMHD} simulations of a \ac{NS} merger with sufficiently high seed
magnetic fields $(10^{15}\, {\rm G})$ to be able to resolve the \ac{MRI}
in the merger remnant and reported averaged $\alpha$ values for
different rest-mass density shells. Combining their estimate of $\alpha$
with values of $c_s$ and $\Omega$ estimated from a simulation performed
with $\ell_{\rm mix} = 0$ we are able to estimate $\ell_{\rm mix}$ as a
function of the rest mass density (Fig.~\ref{fig:lmix_fit}). We find
that the mixing length is well fitted by the expression
\begin{equation}
  \ell_{\rm mix} = \begin{cases}
    a\, \xi\, \exp\Big(-|b\,\xi|^{5/2}\Big)\ [{\rm m}]\,, &\textrm{if }
    \xi > 0\,, \\
    0\,, & \textrm{otherwise}\,,
  \end{cases}
\end{equation}
where 
\begin{equation}
  \xi = \log_{10} \left( \frac{m_b (n_p + n_m)}{\rho^*} \right)\,,
\end{equation}
$m_b$ is the atomic mass unit in grams, $a = 22.31984$, $b =
-0.4253832$, and $\rho^* = 1.966769 \times 10^9\ {\rm g}\ {\rm
cm}^{-3}$. This fit and the constant values of $\ell_{\rm mix}$ used in
our previous studies, are shown in Fig.~\ref{fig:lmix_fit}. 

Our analysis reveals that $\ell_{\rm mix}$ is relatively small even for
the highly-magnetized binary considered by Kiuchi et
al.~\cite{Kiuchi:2017zzg}. The peak value of $\ell_{\rm mix}$ estimated
from the \ac{GRMHD} simulations is remarkably close to the analytic
prediction given by Eq.~(\ref{eq:lambdamri}). We also find that the
turbulence weakens at high densities inside the massive NS
(MNS)\acused{MNS} product of the merger. This is expected, because the
angular velocity deep inside the remnant grows with radius stabilizing
the flow against the \ac{MRI} \cite{Shibata:2005ss, Kastaun:2016yaf,
Hanauske:2016gia, Ciolfi:2017uak, Radice:2017zta}. On the other hand,
the drop of $\ell_{\rm mix}$ at low density is an artifact of our
fitting procedure. Since Kiuchi et al.~\cite{Kiuchi:2017zzg} do not
provide the value of $\alpha$ for densities below $10^{10}\ {\rm g}\
{\rm cm}^{-3}$ we perform a log-linear extrapolation of $\alpha$ to
lower density which results in $\alpha$ becoming zero at the density
$\rho^*$. That said, the value of $\ell_{\rm mix}$ at those densities is
inconsequential for our simulations, because the orbital period for the
part of the disk with density $\rho^*$ is comparable to the total
postmerger simulation time. Overall, we find that turbulence is
strongest in the mantle of the \ac{MNS} and in the inner part of the
disk, at densities between a few times $10^9\ {\rm g}\ {\rm cm}^{-3}$ to
$10^{13}\ {\rm g}\ {\rm cm}^{-3}$.

\subsection{Models and Simulation Setup}
We consider a \ac{BNS} system with component masses (at infinite
separation) $M_A = M_B = 1.35\ M_\odot$. The LS220 \ac{EOS}
\cite{Lattimer:1991nc} is used to describe the nuclear matter. The
initial data is constructed with the \texttt{Lorene}
code~\cite{lorenecode}, while the evolution is performed with
\texttt{WhiskyTHC} using the setup discussed in
Ref.~\cite{Radice:2018pdn}. We simulate the same binary multiple times:
once with the calibrated $\ell_{\rm mix}$ from Sec.~\ref{sec:grles}, and
then with fixed constant values for $\ell_{\rm mix}$: $0$, 5~m, 25~m,
and 50~m. Additionally, each configuration is run twice: with and
without the inclusion of neutrino reabsorption in the simulations.
Neutrino cooling is instead always included. The resolution in the
finest refinement level of the grid, which covers the \acp{NS} during
the inspiral and the \ac{MNS} after merger, is of $185\ {\rm m}$.
Finally, to quantify finite-resolution effects we also rerun the
simulations with no neutrino reabsorption also at the lower resolution
of $246\ {\rm m}$. The results presented here are thus based on a total
of 15 simulations for a total cost of about 3M CPU hours. The
simulations with constant $\ell_{\rm mix}$ were already
presented\footnote{However, we rerun the $\ell_{\rm mix} = 0$ simulation
with neutrino reabsorption, which we now continue for a longer time
after merger than in our previous work.} in Refs.~\cite{Radice:2017zta,
Radice:2018pdn, Radice:2018ghv}.  In Ref.~\cite{Mosta:2020hlh}
postmerger profiles from the $\ell_{\rm mix} = 0$ no neutrino
reabsorption binary was mapped into a high-resolution grid and simulated
with the inclusion of a magnetic field. The simulations with calibrated
turbulence model are new. For clarity, we only include the
high-resolution simulations in the figures.  If not otherwise specified,
the figures refer to the simulations that included both neutrino
emission and neutrino reabsorption. The low-resolution data follows the
same qualitative trends, although there are quantitative differences.

\section{Results}
\label{sec:results}

\subsection{Qualitative Dynamics}

\begin{figure}
  \centering
  \includegraphics[width=5in]{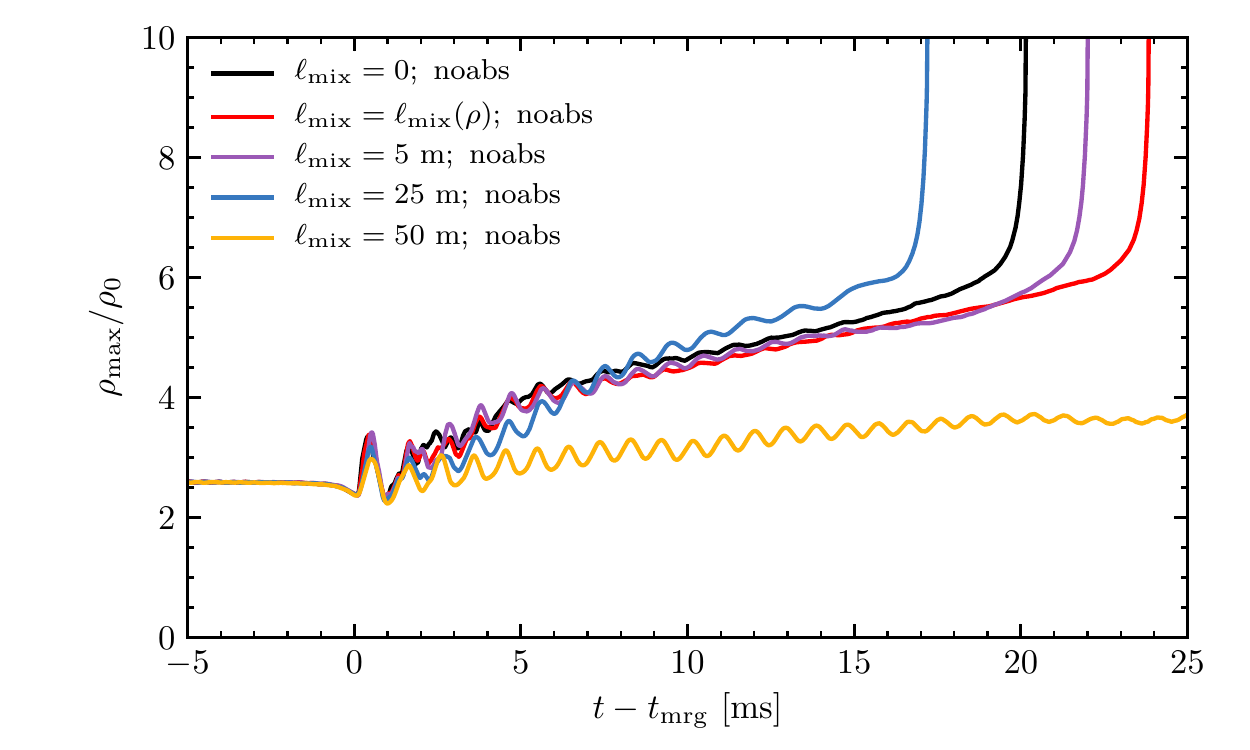}
  \caption{Maximum density evolution for the models computed without
  neutrino heating. We use nuclear saturation density ($\rho_0 =
  2.7\times 10^{14}\ {\rm g}\ {\rm cm}^{-3}$) as density scale. The
  inclusion of turbulent viscosity can drastically alter the lifetime of
  the MNS.}
  \label{fig:rho_max}
\end{figure}

We refer to Refs.~\cite{Bernuzzi:2015opx, Radice:2018pdn} for a detailed
description of the qualitative evolution of the binary considered in
this work. Here, we only mention that the simulations span the last
${\sim}3{-}4$ orbits prior to merger and continue for $20{-}25\ {\rm
ms}$ afterwards. The merger produces a \ac{MNS} remnant that collapses
to \ac{BH} surrounded by a massive accretion torus, typically within the
simulation time. Notable exceptions are the simulations with $\ell_{\rm
mix} = 50\ {\rm m}$ for which collapse appears to be significantly
delayed by viscosity, as we reported in Ref.~\cite{Radice:2017zta}.

The evolution of the maximum rest-mass density for the 5 binaries that
did not include neutrino heating is shown in Fig.~\ref{fig:rho_max}. As
previously reported in Ref.~\cite{Radice:2017zta}, we find that the
turbulence on the lifetime of the remnant is non monotonic. This is due
to the complex interplay between angular momentum transport and
suppression of angular momentum losses to \acp{GW} operated by the
turbulence \cite{Radice:2017zta}. However, only the simulation with the
largest mixing length ($\ell_{\rm mix} = 50\ {\rm m}$), corresponding to
very efficient turbulent transport, shows truly significant differences
in the contraction rate and in the lifetime of the remnant when compared
to the baseline model with $\ell_{\rm mix} = 0$. In the other cases the
changes are quantitative rather than qualitative. This is not
surprising: turbulent viscosity plays a small role in the inner core of
the \ac{MNS} because the velocity gradients are relatively small towards
the center of the \ac{MNS} \cite{Shibata:2005ss, Kastaun:2016yaf,
Hanauske:2016gia}.  These findings are consistent with the results of
the \ac{GRMHD} simulations of the same binary presented in
Ref.~\cite{Mosta:2020hlh}. There it was found that the inclusion or
omission of the magnetic field (and hence of the \ac{MRI}-induced
turbulence) has a modest effect on the collapse time of the remnant,
with difference of the same order as those found in our
Fig.~\ref{fig:rho_max}.

\begin{figure}
  \vspace{-1.5em}
  \includegraphics[width=\textwidth]{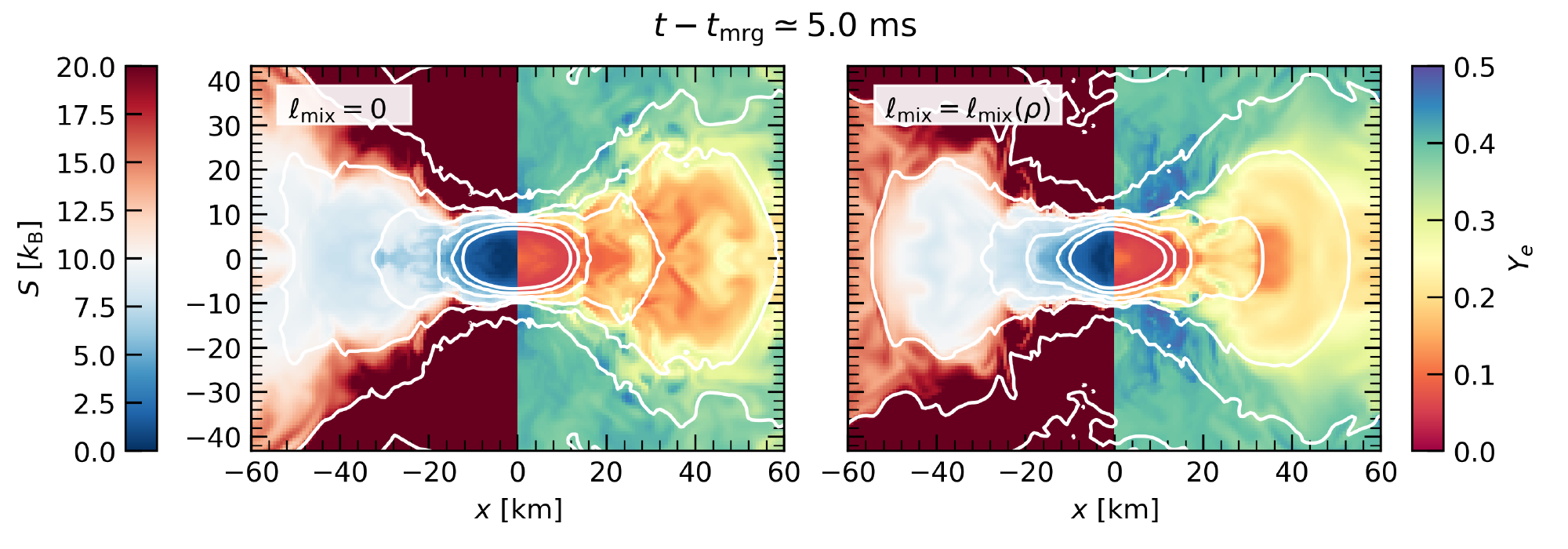}
  \vspace{-0.75em}
  \includegraphics[width=\textwidth]{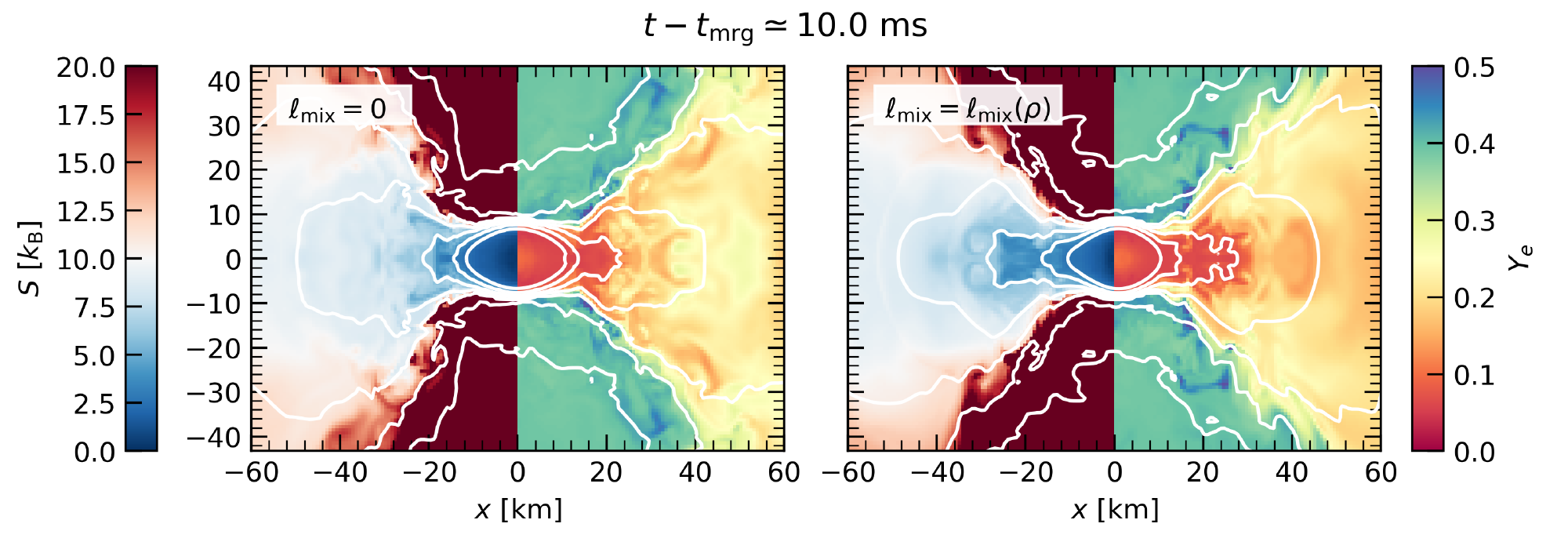}
  \vspace{-0.75em}
  \includegraphics[width=\textwidth]{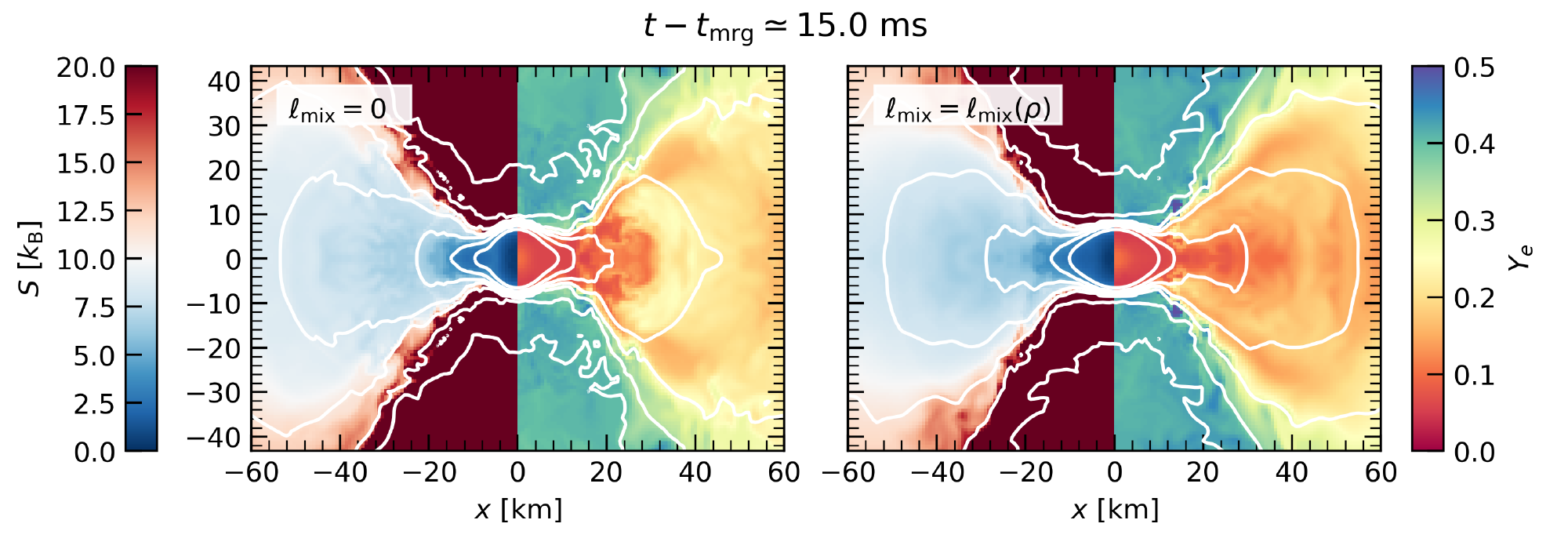}
  \vspace{-0.75em}
  \includegraphics[width=\textwidth]{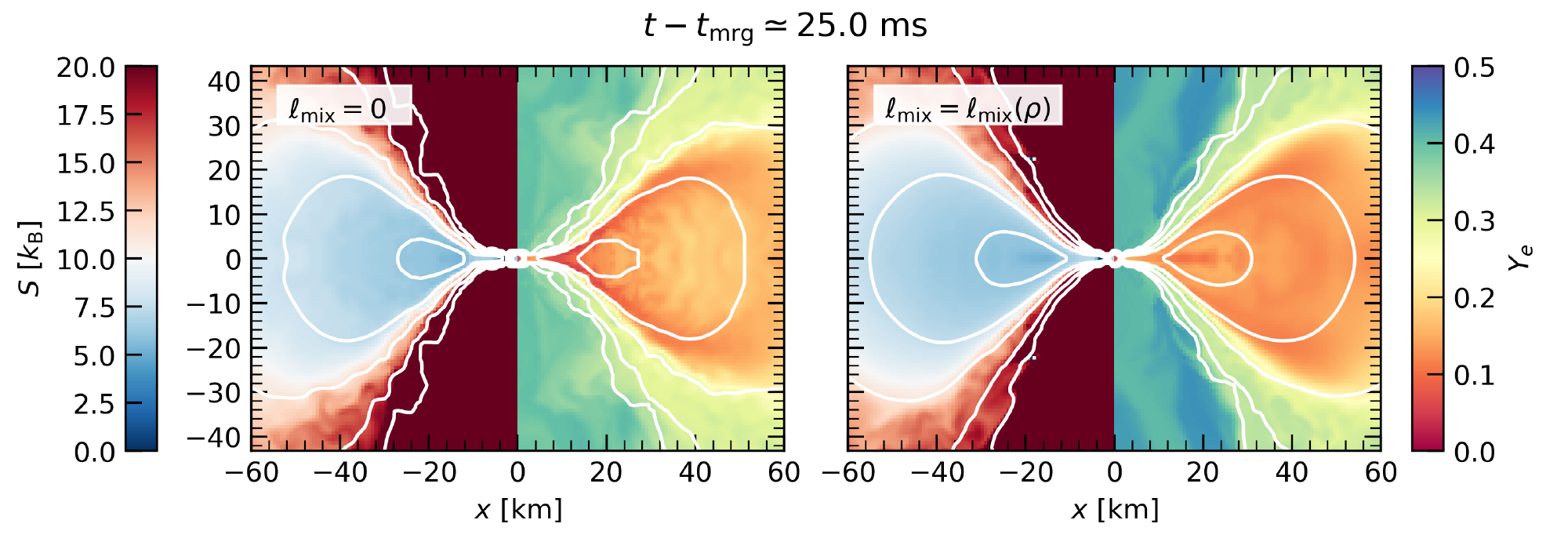}
  \caption{Remnant disk and MNS in the $\ell_{\rm mix} = 0$ and
  $\ell_{\rm mix} = \ell_{\rm mix}(\rho)$ models at four representative
  times. In each panel we show color-coded values of entropy ($x
  < 0$) and electron fraction ($x > 0$) in the meridional plane.. The
  bottom panel shows the disk configuration at the end of the
  simulation, when the MNS has collapsed to BH. The white lines are the
  $10^{8}, 10^{9}, 10^{10}, 10^{11}, 10^{12}, 10^{13},$ and $10^{14}$
  ${\rm g}\ {\rm cm}^{-3}$ isocontours of the rest-mass density.
  Turbulent viscosity mixes material from the mantle of the MNS and the
  inner disks and smooths the structure of the disk.}
  \label{fig:disk2d}
\end{figure}

The effects of turbulent viscosity are more pronounced in the outer
layers of the \ac{MNS} and in the disk, where velocity gradients are
larger. Moreover, $\ell_{\rm mix}$ is maximum in this region according
to the calibrated turbulence model.  The impact of turbulence on the
structure of the disk is shown in Fig.~\ref{fig:disk2d}, where we report
the profiles of entropy, electron fraction, and density for the
$\ell_{\rm mix} = 0$ and $\ell_{\rm mix} = \ell_{\rm mix}(\rho)$ runs
(both with neutrino absorption included). The accretion disk is formed
in the first milliseconds after the merger, as hot material is expelled
from the collisional interface between the \acp{NS}. During this phase,
turbulence dissipation enhances the thermalization of the flow resulting
in the formation of a disk with larger initial entropy and electron
fraction compared to that of the baseline model with $\ell_{\rm mix} =
0$. At later times, turbulence has an opposite effect: turbulent
stresses drive the mixing of the hot material in the inner disk with
fresh low-entropy material from the mantle of the \ac{MNS} lowering
entropy and electron fraction in the inner part of the disk. Over longer
timescales, turbulent angular momentum transport also drives flows of
matter to larger radii. This manifests itself in the increase of the
density in the midplane of the disk and the smoothing of the isodensity
contours in the disk. We remark once again that the internal structure
of the remnant is not visibly affected by the turbulence viscosity (with
the exception of the $\ell_{\rm mix} = 50\ {\rm m}$ run). The apparent
differences in the density isocontours in the \ac{MNS} in
Fig.~\ref{fig:disk2d} arise because the \ac{MNS} has a strong
quadrupolar deformation and the $\ell_{\rm mix} = 0$ and $\ell_{\rm mix}
= \ell_{\rm mix}(\rho)$ simulations are dephased with respect to each
other.

\subsection{Gravitational Waves}

\begin{figure}
  \centering
  \includegraphics[width=5in]{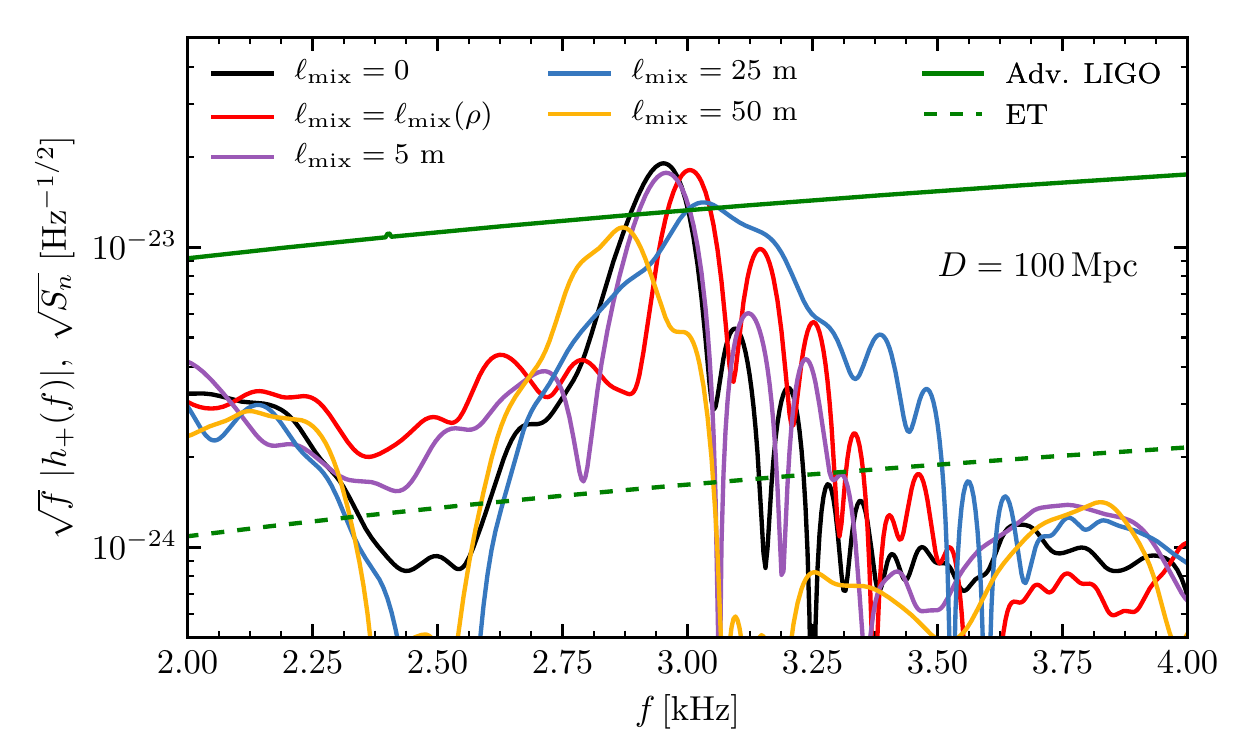}
  \caption{Effective strain the frequency domain for the dominant
  $\ell=2,m=2$ GW multipole for selected models. We window the \ac{GW}
  strain data using an Hann window on the interval $-10\ {\rm ms} < t -
  t_{\rm mrg} < 20\ {\rm ms}$. We show the effective strain for an
  optimally oriented binary at $D = 100\ {\rm Mpc}$. Also shown are the
  design noise curves for Adv.~LIGO, in the high laser power zero
  detuning configuration, and the Einstein Telescope, in the ET-D
  configuration. The effective strain is normalized so that the ratio
  between the signal and the noise curve is equal to the signal to noise
  ratio density in frequency space. Turbulent viscosity results in only
  modest shifts of the dominant postmerger emission frequency. However,
  the subdominant features in the GW spectrum are strongly impacted.}
  \label{fig:spectrum}
\end{figure}

The impact of turbulent viscosity on the \ac{GW} signal is shown in
Fig.~\ref{fig:spectrum}. The most prominent feature of the postmerger
spectrum is a peak at $f_{\rm GW} \simeq 3\ {\rm kHz}$ associated with
the rotational period of the quadrupolarly deformed \ac{MNS}
\cite{Bauswein:2011tp, Hotokezaka:2013iia, Takami:2014zpa,
Bernuzzi:2015rla}. We find this feature to be robust against turbulence:
deviations in the peak frequency with $\ell_{\rm mix}$ are typically
small and of the same order of the nominal uncertainty of the Fourier
transform of the time domain data. Thus, our results confirm that the
measurement of the postmerger peak frequency is a promising avenue to
constrain the \ac{EOS} of dense matter using 3rd generation detectors,
\eg, Ref.~\cite{Breschi:2019srl}. Models with turbulent viscosities
larger than those of our calibrated model, \ie, $\ell_{\rm mix} = 25\
{\rm m}$ and $\ell_{\rm mix} = 50\ {\rm m}$, also show an overall
decrease in the power of the \ac{GW} signal, suggesting that turbulence
might suppress \ac{GW} emission, as also found in
Refs.~\cite{Radice:2017zta, Shibata:2017xht}. That said, significant
reduction in the \ac{GW} signal as found by Ref.~\cite{Shibata:2017xht}
would require turbulent viscosities significantly larger than those
estimated from \ac{GRMHD} simulations \cite{Kiuchi:2017zzg}.

Turbulent angular momentum transport instead has a significant impact on
the secondary features of the postmerger \ac{GW} spectrum. In
particular, we find that turbulence can shift and amplify secondary
peaks in the spectrum that are formed at early time after merger
\cite{Rezzolla:2016nxn}. Such features in the spectrum are analogous to
those due to first order phase transitions \cite{Weih:2019xvw}.
Consequently, we caution that the search for new physics in the
postmerger signal must account for the uncertainties related to the
development of turbulence in the \ac{MNS}. Follow up studies are
necessary to precisely quantify them.

\subsection{Outflows}

\begin{figure}
  \centering
  \includegraphics[width=5in]{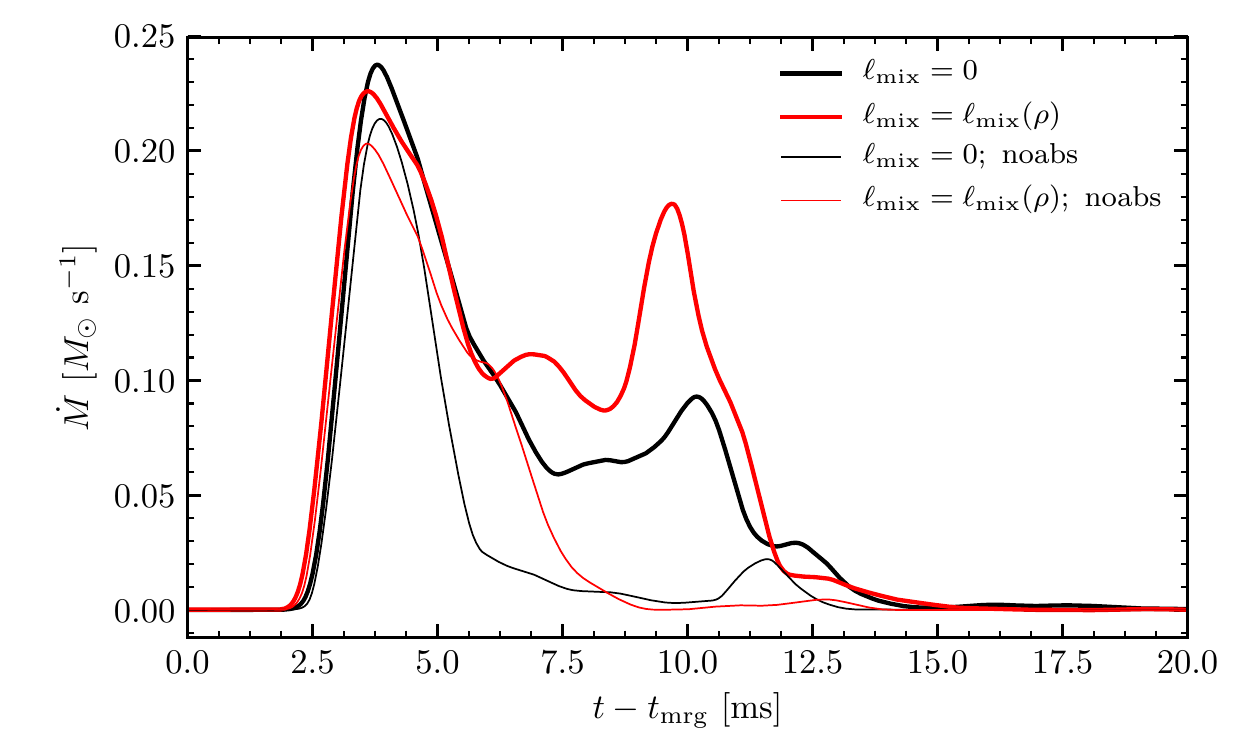} 
  \caption{Outflow rate for the baseline and the calibrated turbulence
  models. The thin lines denote simulations that did not include
  neutrino reabsorption. For clarity we smooth the data using a rolling
  average with amplitude $0.5\ {\rm ms}$. Turbulent dissipation has a
  modest impact on the dynamical ejecta mass and is subdominant in
  comparison to neutrino heating.}
  \label{fig:outflow_lmix}
\end{figure}

Tidal interaction between the stars prior to merger and shocks after
merger drive the ejection of neutron rich matter as the \acp{NS}
coalesce \cite{Shibata:2019wef, Radice:2020ddv}. In
Ref.~\cite{Radice:2018ghv} we studied the impact of turbulent viscosity
on the dynamical ejection of matter during \ac{BNS} mergers.  We found
that turbulence can boost the ejection of matter in asymmetric binaries,
but has only a small impact on the mass ejection from comparable mass
binaries, such as the system considered in this study.  Here, we confirm
that these results hold also when using a calibrated turbulent viscosity
model. In Fig.~\ref{fig:outflow_lmix} we show the outflow rate of
unbound matter (with $u_t \leq -1$) across a coordinate sphere with
radius $R = 200\, G/c^2\, M_\odot \simeq 295\, {\rm km}$.  The
differences in the overall ejecta mass are not large, considering the
large numerical uncertainties associated with the calculation of the
ejecta \cite{Radice:2018pdn}.  However, there is a clear trend in the
outflow data. On the one hand, turbulent dissipation does not effect the
outflow rate at early time, when fast material accelerated when the
remnant rebounds reaches the detection sphere. On the other hand, the
$\ell_{\rm mix} > 0$ runs have significantly larger outflow rates at
later times, when the slower part of the dynamical ejecta crosses the
detection sphere. This is visible for the $\ell_{\rm mix} = \ell_{\rm
mix}(\rho)$ simulations in Fig.~\ref{fig:outflow_lmix}.  Simulations
with other values of $\ell_{\rm mix}$ follow the same trend, but are not
included to avoid cluttering the figure. This late time boost of the
outflow rate is likely the result of the increased thermalization of the
flow due to turbulent viscosity and the resulting larger pressure
gradients. 

The effect of turbulent viscosity is, however, subdominant compared to
the influence of neutrino reabsorption, which are found to play a key
role in driving the ``second wave'' of the dynamical ejecta.  Indeed, as
shown in Fig.~\ref{fig:outflow_lmix}, this part of the outflow is almost
completely absent when neutrino reabsorption is switched off in the
simulations. This second component of the dynamical ejecta is
predominantly constitute of neutrino-irradiated material at high
latitudes above the merger remnant.

\begin{figure}
  \centering
  \includegraphics[width=5in]{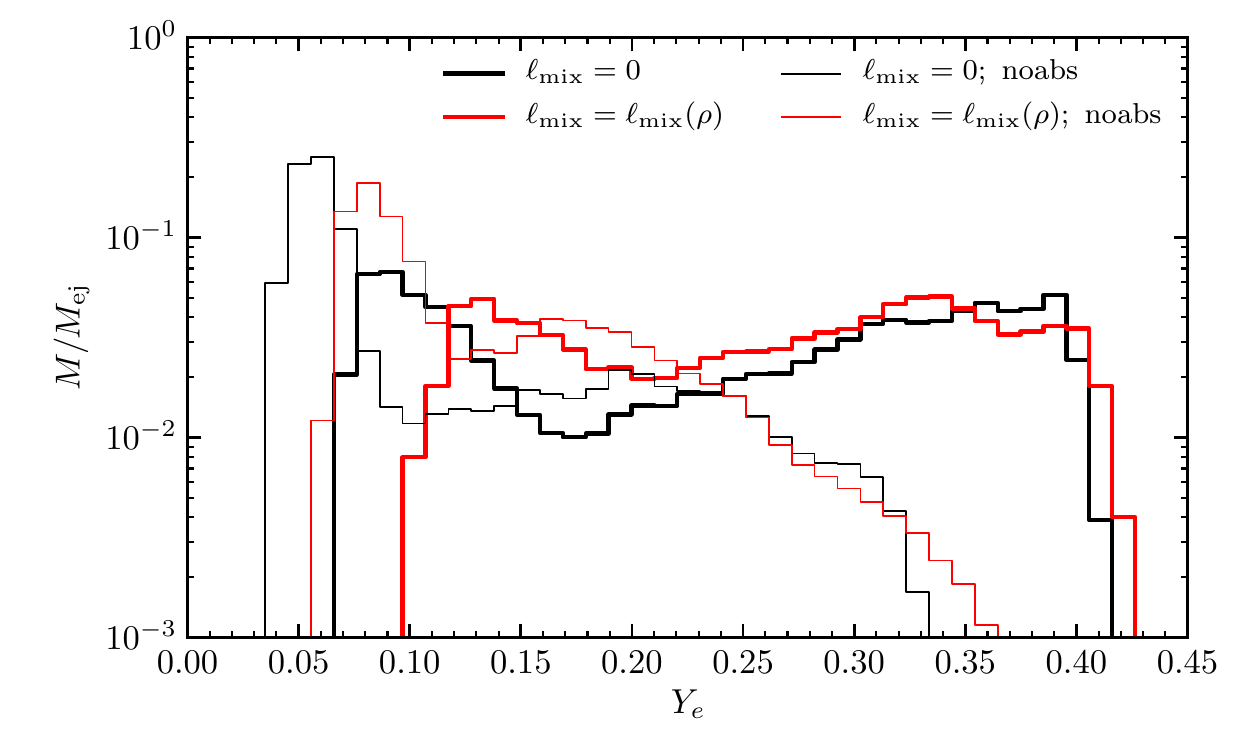} 
  \caption{Histograms of the composition of the outflows from selected
  models. The thin lines denote simulations that did not include
  neutrino reabsorption. Turbulent viscosity and dissipation tend to
  increase the average electron fraction of the ejecta. However, this
  effect is subdominant compared to the effect of neutrino reabsorption.}
  \label{fig:ejecta_ye}
\end{figure}

Both viscosity and neutrino reabsorption play an important role in the
determination of the electron fraction $Y_e$ of the ejecta
\cite{Sekiguchi:2016bjd, Foucart:2015gaa, Radice:2016dwd,
Foucart:2016rxm, Radice:2018pdn}, as shown in Fig.~\ref{fig:ejecta_ye}.
This quantity is of particular interest since it most directly
determines the outcomes of the r-process nucleosynthesis for the
thermodynamic conditions typical of \ac{NS} merger ejecta
\cite{Lippuner:2015gwa}. Turbulent dissipation increases the
temperatures in the outflows activating pair capture processes which
drive $Y_e$ to higher values through the reaction $e^+ + n \rightarrow p
+ \bar{\nu}_e$. This effect is particularly pronounced when considering
the tidal tail, which corresponds to the low-$Y_e$ peak in the outflow
distribution. Neutrino absorption has an even larger effect via the
reaction $\nu_e + n \rightarrow p + e^-$. Neutrinos generate relatively
high $Y_e$ ejecta, especially at high latitudes \cite{Radice:2016dwd}.
They also effect the tidal tail, which is irradiated after colliding
with the faster shock-accelerated outflows that are launched after the
merger \cite{Radice:2018pdn}.

Turbulent angular momentum transport in the remnant accretion disk and
neutrino heating are expected to power additional outflows on a
timescale of a few seconds \cite{Metzger:2008av, Metzger:2008jt,
2009ApJ...699L..93L, Fernandez:2013tya, Just:2014fka, Metzger:2014ila,
Perego:2014fma, Martin:2015hxa, Fujibayashi:2017puw, Siegel:2017jug,
Fernandez:2018kax, Nedora:2019jhl, Fujibayashi:2020qda}. These secular
ejecta are expected to dominate the overall nucleosynthesis and
electromagnetic signal from \ac{BNS} mergers \cite{Radice:2018pdn,
Siegel:2019mlp}. We considered the role of turbulent viscosity on the
long term mass ejection in Ref.~\cite{Nedora:2019jhl}. However, due to
the large computational costs, we could not systematically vary
$\ell_{\rm mix}$ in Ref.~\cite{Nedora:2019jhl}. Unfortunately, the
simulations we present here do not span a sufficiently long time to be
able to study the secular ejecta, so we leave this investigation to
future work.

\section{Discussion}
\label{sec:conclusions}
MHD instabilities active in \ac{BNS} merger remnants drive turbulence at
many different scales \cite{Kiuchi:2015sga, Kiuchi:2017zzg}. Turbulence
can generate large scale magnetic fields with potentially dramatic
consequences for the evolution of the \ac{MNS} remnants and its \ac{EM}
emissions \cite{Siegel:2013nrw, Ciolfi:2019fie, Mosta:2020hlh}. Due to
the large Reynolds number in the flow, directly capturing all scales of
the turbulent flow in the \ac{MNS} is beyond the reach of numerical
simulations for the foreseeable future. 

In Ref.~\cite{Radice:2017zta} we proposed a scheme to include subgrid
scale turbulence effects into global simulations and study the
associated uncertainties on the multimessenger signatures of \ac{BNS}
mergers. Our method extends the \ac{LES} methodology to \ac{GR}. We
derived evolution equations for coarse-grained fluid number,
momentum, and energy densities. These equations are exact, but are not
closed, so a closure must be provided.  This closure represents the
effect of small scale (subgrid) turbulence on the evolution of the large
scale quantities. Here, we proposed to use a closure based on the
classical turbulent viscosity ansatz \cite{Smagorinsky:1963a}, which we
calibrated against very-high-resolution \ac{GRMHD} simulations from
Ref.~\cite{Kiuchi:2017zzg}. We performed \ac{BNS} merger simulations
with microphysics and neutrinos using the newly proposed turbulence
model. We showed that our scheme is robust and gives sensible results.
We compared simulations performed with the newly calibrated scheme with
simulations performed either with no subgrid model, or with a simpler
scheme in which turbulent viscosity is assumed to be a constant fixed on
the basis of dimensional analysis arguments.

Our results show that subgrid turbulence has a quantitative impact on
the evolution of the remnant. Turbulence can affect the lifetime of the
remnant, although large differences in the postmerger evolution are only
found for values of the turbulent viscosity that are significantly
larger than those found in \ac{GRMHD} simulations. The peak frequency of
the postmerger \ac{GW} spectrum is found to be insensitive to
turbulence, but secondary features in the spectrum and the overall
luminosity in the postmerger are affected by turbulent dissipation.
Turbulent angular momentum transport and dissipation alter the structure
and composition of the remnant disk and the amount and composition of
the dynamical ejecta, potentially impacting r-process nucleosynthesis
yields and \ac{EM} counterparts. However, turbulence is found to be
subdominant when compared to neutrino effects.

More simulations are needed to establish the degree to which systematic
uncertainties due to turbulence will limit our ability to search for new
physics, such as phase transitions, in multimessenger observations of
\ac{BNS} mergers. The method we propose here can be and has been
extended to the full-\ac{GRMHD} equations \cite{Carrasco:2019uzl,
Vigano:2020ouc}. Including \ac{GRMHD} in our simulations is crucial to
capture also the large scale effects due magnetic fields, such as jet
launching \cite{Ruiz:2016rai, Mosta:2020hlh}, which are presently not
included. In parallel, local high-resolution simulations should be
performed to develop and calibrate turbulence models. These are all
objectives of our future work.

\vspace{6pt}

\funding{
This research used resources of the National Energy Research Scientific
Computing Center, a DOE Office of Science User Facility supported by the
Office of Science of the U.S.~Department of Energy under Contract
No.~DE-AC02-05CH11231. Simulations were also performed on the
supercomputers Comet and Stampede (NSF XSEDE allocation TG-PHY160025),
on NSF/NCSA Blue Waters (NSF AWD-1811236), and on the Advanced Computer
Infrastructure (ACI) of the Institute for Computational and Data Science
(ICDS) at the Pennsylvania State University.
}

\acknowledgments{
It is a pleasure to thank S.~Bernuzzi for discussions that motivated
this work, A.~Prakash for carefully proofreading the manuscript,
L.~Weih and L.~Rezzolla for discussions, and S.~Hild for the ET-D noise
curve data.
}

\conflictsofinterest{The author declares no conflict of interest.}

\reftitle{References}

\externalbibliography{yes}
\bibliography{references}

\acrodef{ADM}{Arnowitt-Deser-Misner}
\acrodef{AMR}{adaptive mesh-refinement}
\acrodef{BH}{black hole}
\acrodef{BBH}{binary black-hole}
\acrodef{BHNS}{black-hole neutron-star}
\acrodef{BNS}{binary neutron star}
\acrodef{CCSN}{core-collapse supernova}
\acrodefplural{CCSN}[CCSNe]{core-collapse supernovae}
\acrodef{CMA}{consistent multi-fluid advection}
\acrodef{DG}{discontinuous Galerkin}
\acrodef{HMNS}{hypermassive neutron star}
\acrodef{ECSS}{Extended Collaborative Support Service}
\acrodef{EM}{electromagnetic}
\acrodef{ET}{Einstein Telescope}
\acrodef{EOB}{effective-one-body}
\acrodef{EOS}{equation of state}

\acrodef{FF}{fitting factor}
\acrodef{GR}{general relativity}
\acrodef{GRLES}{general-relativistic large-eddy simulation}
\acrodef{GRHD}{general-relativistic hydrodynamics}
\acrodef{GRMHD}{general-relativistic magnetohydrodynamics}
\acrodef{GW}{gravitational wave}
\acrodef{KH}{Kelvin-Helmholtz}
\acrodef{ILES}{implicit large-eddy simulations}
\acrodef{LIA}{linear interaction analysis}
\acrodef{LES}{large-eddy simulation}
\acrodef{LVC}{LIGO/Virgo collaboration}

\acrodef{MHD}{magnetohydrodynamics}
\acrodef{MNS}{massive neutron star}
\acrodef{MRI}{magnetorotational instability}
\acrodef{NR}{numerical relativity}
\acrodef{NS}{neutron star}
\acrodef{PNS}{protoneutron star}
\acrodef{SASI}{standing accretion shock instability}
\acrodef{SGRB}{short $\gamma$-ray burst}
\acrodef{SN}{supernova}
\acrodefplural{SN}[SNe]{supernovae}
\acrodef{SNR}{signal-to-noise ratio}

\end{document}